\documentclass[twocolumn]{aastex631}

\usepackage{amsmath}
\usepackage{subfigure}
\definecolor{blue-violet}{rgb}{0.30, 0.1, 0.89}

\begin{document}

\title{The Intrabinary Shock and Companion Star of Redback Pulsar J2215+5135}

\author[0000-0002-9545-7286]{Andrew G. Sullivan}
\affiliation{Kavli Institute for Particle Astrophysics and Cosmology, Department of Physics, Stanford University, Stanford, CA 94305, USA}

\author[0000-0001-6711-3286]{Roger W. Romani}
\affiliation{Kavli Institute for Particle Astrophysics and Cosmology, Department of Physics,
Stanford University, Stanford, CA 94305, USA}



\begin{abstract}
PSR J2215+5135 (J2215) is a `redback' spider pulsar, where the intrabinary shock (IBS) wraps around the pulsar rather than the stellar-mass companion. Spider orbital light curves are modulated, dominated by their binary companion thermal emission in the optical bands and by IBS synchrotron emission in the X-rays. We report on new {\it XMM}-Newton X-ray and U-band observations of J2215. We produce orbital light curves and use them to model the system properties.  Our best-fit optical light model gives a  neutron star mass $M_{NS}=1.98\pm0.08$\,M$_\odot$, lower than previously reported. However, uncertainty in the stellar atmosphere metallicity, a parameter to which J2215 is unusually sensitive, requires us to consider an acceptable systematic plus statistical range of $M_{NS}\sim1.85-2.3$\,M$_\odot$. From the X-ray analysis, we find that the IBS wraps around the pulsar, but with a pulsar wind to companion wind momentum ratio unusually close to unity, implying a flatter IBS geometry than seen in other spiders. Estimating the companion wind momentum and speed from the X-ray light curve, we find a companion mass-loss rate of ${\dot M}_c\gtrsim10^{-10}$ M$_\odot$\,yr$^{-1}$, so that J2215 may become an isolated millisecond pulsar in $\sim 1$ Gyr. Our X-ray analyses place constraints on the magnetization and particle density of the pulsar wind and support models of magnetic reconnection and particle acceleration in the highly magnetized relativistic IBS.
\end{abstract}



\section{Introduction} \label{sec:intro}
Spider pulsars found in {compact} binaries with a low-mass companion star have orbital periods $P_b\lesssim1$ day. The companion mass determines whether the spider falls into the black widow ($M_c< 0.1$ M$_\odot$) or redback ($M_c\approx 0.1-0.4$ M$_\odot$) category. In these systems, pulsar gamma-ray emission and the relativistic particles from the pulsar wind irradiate the companion, consequently driving off a massive stellar wind \citep{1988Natur.334..225K,1988Natur.334..684V, 2013IAUS..291..127R, 2019Galax...7...93H}.  The pulsar wind and companion wind collide to form an intrabinary shock (IBS). In redbacks {which typically have higher stellar wind momentum fluxes}, the companion wind dominates the pulsar wind so the IBS wraps around the pulsar, while in black widows {with lower stellar wind momentum fluxes}, the IBS wraps around the companion, as the pulsar wind dominates the stellar wind \citep{2016ApJ...828....7R, 2017ApJ...839...80W, 2019ApJ...879...73K}. 

X-ray observations {may} probe the IBS while optical observations typically reveal the companion heating. Particles in the shocked pulsar wind accelerate and emit synchrotron X-rays \citep{2019ApJ...879...73K, 2021ApJ...917L..13K}. The light curves are characterized by two caustic peaks per orbital phase, associated with the beamed emission from relativistic particles traveling tangent to the observing line of sight. Thermal emission from the companion is generally boosted by reprocessed pulsar gamma-rays on the `day' side of the companion and sometimes by precipitating IBS particles reprocessed in surface hot spots.
Consequently, X-ray light curve and spectral analyses reveal the IBS structure and particle acceleration mechanisms, while optical analyses examine heating of the companion surface.

PSR J2215+5135 (J2215 hereafter) is a redback millisecond pulsar (MSP) with spin period $P_s=2.61$ ms and spin-down power $\dot{E}=5\times10^{34}$  $I_{45}$ erg s$^{-1}$ in a $P_b=4.14$ hr orbit with a $M\approx0.3$ M$_\odot$ companion \citep{2011AIPC.1357...40H}. The source has dispersion measure of 69.2 pc cm$^{-3}$ \citep{2011AIPC.1357...40H}, which corresponds to an estimated distance $\sim3$ kpc \citep[e.g.]{2017ApJ...835...29Y}. First detected as an unidentified gamma-ray source by the {\it Fermi} Large Area Telescope ({\it Fermi}-LAT) \citep{2009ApJ...697.1071A} and discovered by 350 MHz Green Bank Telescope followup observations \citep{2012arXiv1205.3089R, 2011AIPC.1357...40H}, J2215 has been studied optically \citep{2013ApJ...769..108B, 2014ApJ...793...78S, 2015ApJ...809L..10R, 2018ApJ...859...54L}, in radio \citep{2011AIPC.1357...40H, 2016MNRAS.459.2681B},  in gamma-rays \citep{2015ApJ...809L..10R}, and briefly in X-rays \citep{2014ApJ...783...69G, 2014ApJ...795...72L} prior to this work.

Earlier optical light curve modeling suggests that J2215 contains a particularly massive neutron star $\gtrsim 2$ M$_\odot$ \citep[][hereafter KR20]{2018ApJ...859...54L, 2020ApJ...892..101K}. KR20 estimate a neutron star mass of $M_{NS}=2.24\pm0.09$ M$_\odot$ using a companion surface optical model that contains a magnetic pole hot spot. Such a massive neutron star would constrain properties of the dense matter equation of state \citep{2007PhR...442..109L, 2013ApJ...765L...5S, 2023PhRvD.108i4014B}. Previous modeling is sensitive to the poorly constrained companion hot spot, adding uncertainty to the binary inclination $i$ and, by extension, $M_{NS}$ estimates. Unusually, the previous best-fit companion model has sub-solar metallicity with $\log Z =-1$. Also, older X-ray observations, which
suggested widely separated light curve peaks and a flat IBS, relied on an {\it XMM}-Newton exposure with strong background flaring and only two binary orbits. Thus, we conduct more detailed optical and X-ray study to better understand the robustness and systematics of the companion heating model and probe the IBS physics.

In this paper, we present the results of light curve analyses of new {\it XMM}-Newton J2215 X-ray and U band observations. In sec.\,\ref{sec:Observation}, we summarize the new data, presenting the X-ray and U light curves. In sec.\,\ref{sec:Spectra}, we discuss our analysis of the collected X-ray spectra. In sec.\,\ref{sec:LCModel}, we show the results of optical companion heating and X-ray IBS light curve analyses. We discuss our results and their implications in sec.\,\ref{sec:conclusion}. 

\section{XMM-Newton Observations}
\label{sec:Observation}
The {\it XMM}-Newton Observatory \citep{2001A&A...365L...1J} performed three new observations of J2215, the first two on 2022 June 8-11 (ObsIDs 0900770101 and 0900770201) and the third on 2022 December 1 (ObsID 0900770301) lasting a total of 120.7 ks. These observations provided both X-ray and U band data, which are presented here for the first time.  We supplement these new observations with an archival 54.9 ks {\it XMM}-Newton X-ray observation (ObsID 0783530301). We perform data {reduction} using tools of the XMM-Newton Scientific Analysis System (SAS) \citep{2001A&A...365L...1J}. 
\subsection{X-rays}
We process the EPIC-PN (PN) and EPIC-MOS (MOS) X-ray data using the SAS tools epproc and emproc. We perform standard barycenter corrections using the SAS barycen tool and filter out background flaring events. After this processing, the total exposure times for PN and the MOS cameras are 100.9 ks and 128.5 ks across 14 orbits. We extract the source from a circular region with radius $\sim 30"$ and select circular background regions on the same detector chip as the source with radii 4-5 times larger than the source aperture. We obtain background subtracted X-ray light curves for the three detectors from the SAS tool epiclccorr.

For the orbital parameters, we use the {\it Fermi}-LAT third gamma-ray pulsar catalog ephemeris valid from 2008 August 4 until 2019 December 22 for all observations \citep{2023ApJ...958..191S}, as this is the most recent published ephemeris for J2215. The times of ascending node (TASC) deviate from the mean by less than 0.003 in phase for the entire ephemeris range, so we expect no significant shift in the TASC at our epoch (although 1.5-2 yr outside the ephemeris range). The observed phase of the Optical Monitor (OM) U light curve, to be discussed in the following subsection, is in good agreement with the ephemeris prediction.

\begin{figure*}
    \centering
    \includegraphics[width=0.9\linewidth]{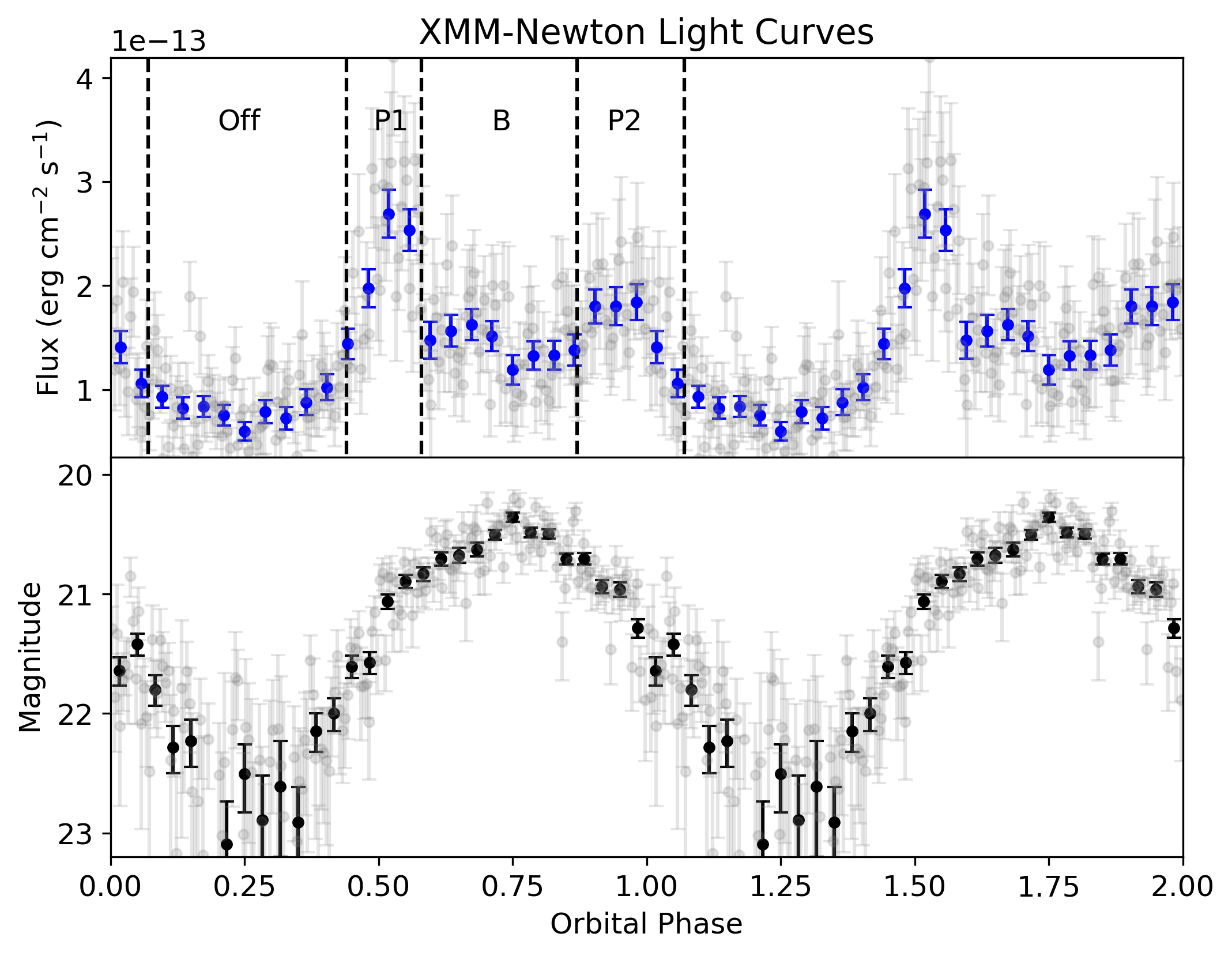}
    \caption{{Binned X-ray and optical orbital light curves of J2215. More-finely binned light curves are shown by faded markers.} (Top) The combined PN and MOS 0.5-10 keV X-ray light curve. We identify four distinct regions for which we separately fit spectra: two peaks, P1 and P2 corresponding to the two phase intervals in which the flow of the IBS shock is tangent to our line of sight, a bridge (B) between the two peaks, and an off region around the companion inferior conjunction where the X-ray emission comes principally from the IBS nose. (Bottom) The OM U filter light curve for J2215 from the new {\it XMM}-Newton observations. }
    \label{fig:XrayLC}
\end{figure*}

We convert the PN and MOS light curves from count rate and count rate error to flux in each detector. We combine the data and distribute it into 26 phase bins. We compute the average flux weighted by the inverse square of the errors in each bin. Flux errors are estimated by the standard error of a variance-weighted sample. 
Our combined 0.5-10 keV X-ray light curve is shown in the top panel of Fig. \ref{fig:XrayLC}.
\subsection{Optical}
\label{sec:opticaldata}
Contemporaneous OM U filter data was taken in both image and fast mode. The fast mode 10$^{\prime\prime}\times10^{\prime\prime}$ images were obtained from a series of exposures over each observation. We use the SAS tool omfchain to obtain U filter light curves. For each exposure, we extract the source from a circular aperture with a radius of 3 detector pixels in the fast mode image and determine the background from the average count rate of the corresponding image mode image with all point sources removed. The extracted count rates are subsequently corrected by the fraction of the point-spread function (PSF) contained in the aperture. After processing, the OM exposure time is 85 ks. We convert the OM U data to flux using the published OM U calibration for A-type stars. We place the data into flux bins by computing the sample mean and standard error. We also use the {\it Fermi}-LAT third gamma-ray pulsar catalog ephemeris for the orbital parameters. The OM U light curve is shown in the bottom panel of Fig. \ref{fig:XrayLC}. 
\section{Spectral Analysis}
\label{sec:Spectra}
\begin{figure}
    \centering
    \includegraphics[width=\columnwidth]{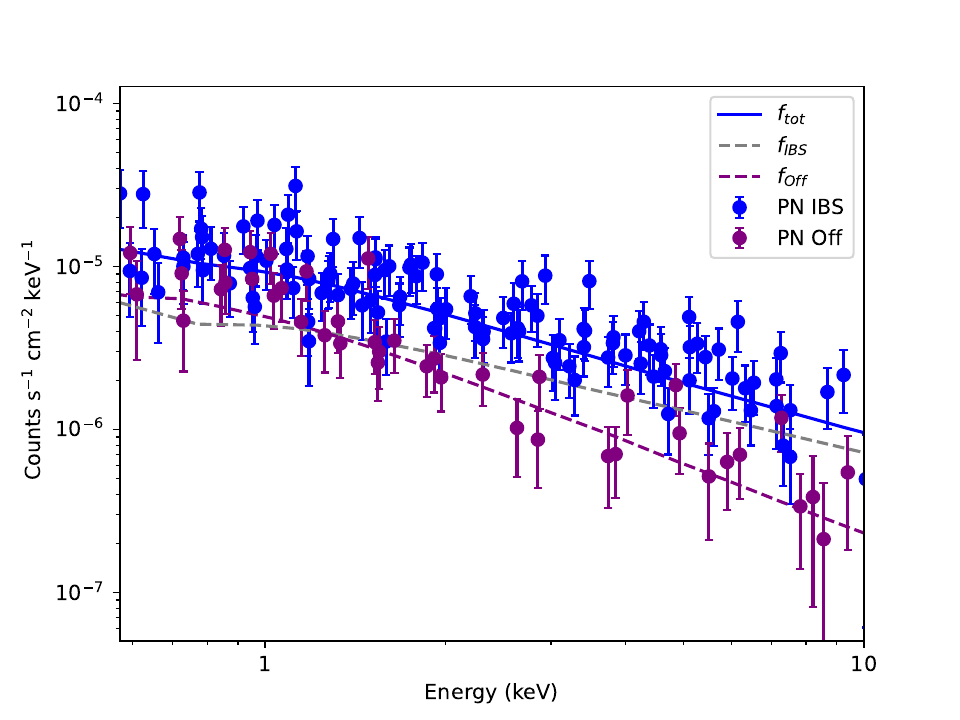}
    \caption{The observed PN X-ray spectra with the equivalent hydrogen column $N_H$ fixed at $0.09\times10^{22}$ cm$^{-2}$. The plotted points are the spectra in the IBS phase region ($0.44<\Phi<1.07$; including P1, P2,  and B) as well as in the off region ($0.07<\Phi<0.44$). The  solid line shows the full model in the IBS phases $f_{tot}$ and the two dashed lines show the absorbed power-law models for the IBS component $f_{IBS}$ with power-law index $\Gamma_{IBS}$ and the phase-independent component $f_{Off}$ with power-law index $\Gamma_0$. Note that in the off region, the only  model component is the absorbed power-law $f_{Off}$.} 
    \label{fig:spectra}
\end{figure}

We divide our X-ray light curve into four phase intervals as labeled in Fig. \ref{fig:XrayLC} and perform spectral analyses of these phase bins. We identify Peak 1 (P1; {$0.44<\Phi<0.58$}) and Peak 2 (P2; {$0.87<\Phi<1.07$}) as the two characteristic IBS X-ray peaks \citep{2016ApJ...828....7R, 2019ApJ...879...73K}. We demarcate a bridge region (B; {$0.58<\Phi<0.87$}) between P1 and P2 and an off region {($0.07<\Phi<0.44$)} corresponding to the companion at inferior conjunction. Here non-IBS fluxes (e.g. from the magnetosphere or a pulsar wind nebula) may dominate, as only weakly beamed X-rays from the IBS nose are expected. We use the CIAO tool Sherpa \citep{2001SPIE.4477...76F, 2007ASPC..376..543D} to fit non-thermal power-law spectral models to the {\it XMM} data with spectral indices $\Gamma_{P1}$, $\Gamma_{P2}$, and $\Gamma_B$. We include an additional phase-independent power-law component throughout the orbit with power-law index $\Gamma_0$ to account for phase-independent emission.
The spectral model in counts s$^{-1}$ cm$^{-2}$ keV$^{-1}$ in each phase region is
\begin{equation}
    f_r(E)=e^{-N_H \sigma(E)}\left(K_{r}E^{-\Gamma_r}+K_{0}E^{-\Gamma_0}\right),
\end{equation}
where $N_H$ is the equivalent hydrogen column, $\sigma(E)$ is the photo-electric cross section, $K_r$ and $K_0$ are power-law normalization factors, and $r=\{P1, P2, B, \textit{Off}\}$ demarcates the orbital phase region. {For our absorption model, we employ the photoelectric cross sections of \cite{1992ApJ...400..699B} and the elemental abundances of \cite{1989GeCoA..53..197A}.} In the off phase, $K_r=0$. 
The results of our fit are shown in Table \ref{table:1}. Because the spectral indices of  P1, P2, and B are not significantly different, we also give the average spectral index across the IBS $\Gamma_{IBS}$. {Fig. \ref{fig:spectra} shows the spectra across the IBS as well as in the off phases as observed by PN.}

\begin{table}
\setlength{\tabcolsep}{4pt}
\begin{tabular}{ccc}
 
 \hline\hline
   &Fixed $N_H$&  Free $N_H$  \\
 \hline
$N_H$ ($10^{22}$ cm$^{-2}$) & 0.09& $0.10 \pm 0.02$ \\
$\Gamma_{P1}$ & $0.99\pm0.11$& $1.00\pm0.11$\\
$\Gamma_{P2}$ & $0.85\pm0.12$&$0.86\pm0.12$ \\
$\Gamma_B$ &  $0.85\pm0.15$&$0.86\pm0.12$\\
$\Gamma_{IBS}$ & $0.94 \pm0.08$ &$0.96\pm0.08$ \\
$\Gamma_0$ & $1.43\pm0.08$&$1.48\pm0.12$
\\
$F_{tot}$ ($10^{-14}$ erg cm$^{-2}$ s$^{-1}$) & $15.1\pm1.6$ & $15.2\pm1.9$\\
$F_{IBS}$ ($10^{-14}$ erg cm$^{-2}$ s$^{-1}$) & $9.9\pm1.6$ & $10.0\pm1.8$\\
$F_0$ ($10^{-14}$ erg cm$^{-2}$ s$^{-1}$) & $5.3\pm0.6$ & $5.2\pm1.0 $\\
$\chi^2$/DoF & 1.03 & 1.03 \\
 \hline\hline
\end{tabular}
\caption{X-ray spectral model fits. We fit a phase-independent power-law with index $\Gamma_0$ throughout the orbit and add additional power-laws with separate spectral indices in the P1, P2, and B phase regions. {The total measured flux in the IBS phase region $0.44<\Phi<1.07$ is $F_{tot}$. The measured flux of the specific IBS component is $F_{IBS}$, while the phase-independent component is $F_0$}. The equivalent hydrogen column $N_H$ is fixed at $0.09\times10^{22}$ cm$^{-2}$, the value inferred from 3D dust maps \citep{2018MNRAS.478..651G}, in the left column, while allowed to vary in the right.}
\label{table:1}
\end{table}
\section{Light Curve Modelling}
\label{sec:LCModel}
\subsection{Optical Models}

Previous optical modeling of J2215 has been conducted by \cite{2018ApJ...859...54L} and KR20 using g'r'i' data from two observing nights at the William Herschel Telescope (WHT). These light curve analyses show strong heating asymmetries on the companion star, suggesting the presence of a hot spot, which could plausibly be a magnetic pole \citep{2017ApJ...845...42S}. KR20 include a hot spot at the coordinates $(\theta_{HS}, \phi_{HS})$ defined from the nose on the stellar surface. At that location, the stellar temperature increases by a factor $(1+A_{HS})$ from $T\approx5600$ K to $T\gtrsim8000$ K with a Gaussian profile over radius $r_{HS}$. Their best-fit model gives $i=68.9^\circ$ and consequently a very high $M_{NS}=2.24$ M$_\odot$. These results, however, are sensitive to the hot spot parameters. Additionally, they were obtained using a simplified gravity darkening model, which does not account for the gravity darkening of the reprocessed pulsar heating luminosity \citep{2011A&A...529A..75C, 2023ApJ...942....6K}. 

We refit the WHT g'r'i' data along with the new OM U data using the ICARUS light curve modeling code \citep{2013ApJ...769..108B} including the extensions described in KR20 and the revised gravity darkening law described in \cite{2023ApJ...942....6K}. The model is very similar to the best-fit hot spot model of KR20; we augment a direct pulsar heating model with a surface hot spot and add a flat spectrum veiling flux $f_2$ to the WHT night 2 data. We adopt an extinction $A_V=0.4$ in these fits, as this matches the fit $N_H$ (Table 1) and is consistent with estimates from current 3-D dust maps \citep{2018MNRAS.478..651G}.  We perform Monte Carlo sampling of the parameters using the library PyMultinest \citep{2008MNRAS.384..449F, 2009MNRAS.398.1601F, 2014A&A...564A.125B, 2019OJAp....2E..10F}, which uses nested sampling to search the parameter space. To predict the flux in each photometric band, we use BT-Settl model atmospheres \citep{2014IAUS..299..271A} available from the Spanish Virtual Observatory ({\tt http://svo2.cab.inta-csic.es}). These model atmosphere spectra are available with a range of metallicities. We have explored models with $\log Z=-1.5$ to $+0.5$; a few models are also available with varying degrees of $\alpha$ process abundance enhancement.

Our fits prefer calibration offsets $\delta m_r\gtrsim0.03$ to both nights in the WHT r' band and $\delta m_U\approx0.5$ to the OM U data. We have attempted to find the origin of this large U offset. As noted in Sec.\,\ref{sec:opticaldata}, we obtain the OM U AB magnitudes by converting from count rate to flux with the SAS-recommended U conversion factor of $1.7 \times 10^{-16}$ erg cm$^{-2}$ \AA$^{-1}$ count$^{-1}$ for A stars. This spectral type is appropriate for the companion near optical maximum. Using archival {\it XMM} OM U exposures in fields covered by the Sloan Digital Sky Survey (SDSS), where catalog $u$ magnitudes are available, we have also investigated the calibration manually. We extracted the count rates in 3-pixel apertures centered on bright stars in two SDSS fields, subtracted the average backgrounds in the images, and scaled the stars' count rates by the calibration aperture correction factor given by SAS. From these data we find a conversion to SDSS u of $2.7 \pm 0.2 \times 10^{-16}$   erg  cm$^{-2}$ \AA$^{-1}$ count$^{-1}$, which would suggest that the U fluxes are brighter by a factor of $\sim1.6$ (or offset by $\sim0.5$ mag) with respect to the our {\it XMM} calibrated flux, comparable to the fit offset.

We have also examined the calibrated 2014 Keck LRIS spectra of \cite{2015ApJ...809L..10R}, whose g'r'i' colors are consistent with the 2014 WHT photometry. We find the Keck fluxes at wavelengths below $4000$\,\r{A} are consistent with the family of BT-Settl model atmosphere spectra \citep{2014IAUS..299..271A} at the day side temperature of the star. A change to the thermal companion flux (unlike the IBS non-thermal flux) seems unlikely and would require a concomitant change in the heating power. Our X-ray flux is consistent with past values and there is no significant change to the (dominant) {\it Fermi}-LAT gamma-ray heating. Consequently, we consider a U calibration offset the most likely explanation. While fitting for $\delta m_U$ slightly decreases the predictive power of our model, the OM U light curve shape still probes the hot spot geometry.

\begin{figure}
    \centering
    \vskip -1.3cm
    \includegraphics[width=1.1\columnwidth]{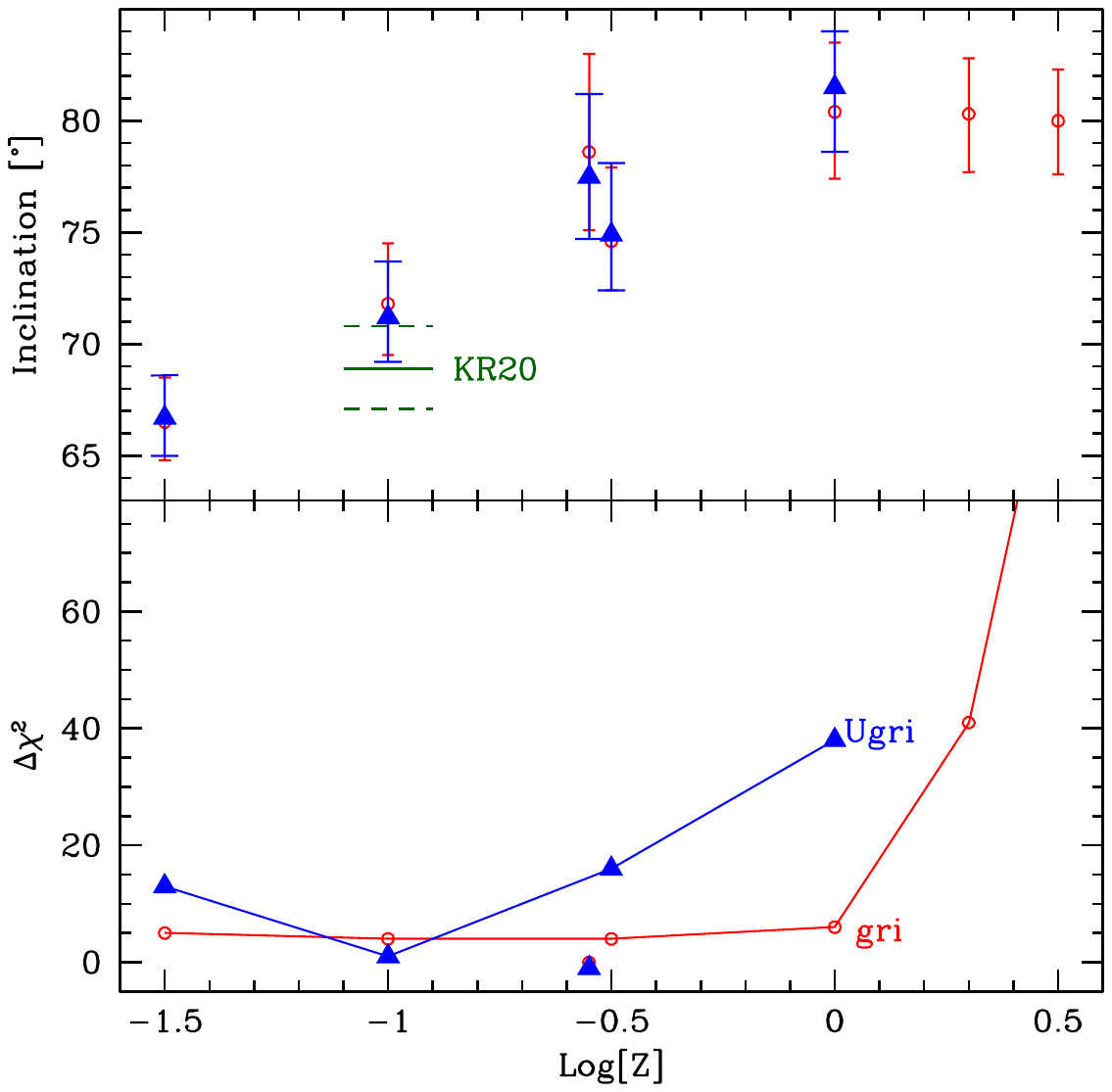}
    \vskip -2.5cm  
    \caption{
    Sensitivity of the light curve fits to $\log Z$. The top panel shows the best-fit inclination, while the lower panel shows the $\Delta \chi^2$ from minimum. Fits of g$^\prime$r$^\prime$i$^\prime$ alone are shown in red while fits of Ug$^\prime$r$^\prime$i$^\prime$ are shown in blue. For g$^\prime$r$^\prime$i$^\prime$ all $\log Z \le 0$  metallicities are acceptable with near-identical $\chi^2$, leaving large $i$ (and mass) uncertainty. Including {\it XMM} U breaks this degeneracy, preferring $\log Z=-1$ (the value used in KR20). The left points of the $\log Z=-0.5$ pair are for a lower $\alpha=0$ element abundance; these represent the global minimum values found in the fits. See the text for details on $\alpha$ abundance.} 
    
    \label{fig:logzvsincl}
\end{figure}

When fitting the g'r'i' light curve data without including U, we find that the best-fit parameters for $i$ and the hot spot are, unusually, sensitive to the metallicity of the companion stellar atmosphere. This happens for J2215 because the location of the hot spot allows it to mimic a substantial portion of the companion nose heating under the corrected gravity darkening law. With only g'r'i' colors, the spot location and temperature are not independently constrained, so there is substantial covariance with $i$ in the fits. Consequently, the subtle metallicity-dependent color differences shift the best-fit $i$ and hot spot parameters. For each atmosphere model, we show the $\Delta \chi^2$ from the best fitting metallicity model as well as the estimated inclination in Fig.\,\ref{fig:logzvsincl}. This hot spot metallicity sensitivity seems to be a peculiarity of J2215's parameters; we do not see such sensitivity in our fits of other objects \citep{2019ApJ...883..108D}. Nevertheless, this remains a possible rare systematic that should be checked in spider model fitting.

The enhancement of $\alpha$-process elements increases as $\log Z$ decreases in the BT-Settl models. For $\log Z\ge 0$, all models have $\alpha=0$, while for
$\log Z\le -1$ the bulk of the models have $\alpha=0.4$. At $\log Z=-0.5$ most models have $\alpha=0.2$, but there are a subset at $\alpha=0$ that span fewer temperatures and surface gravity values.
With the WHT g'r'i' data, we find that super-solar metallicities are unacceptable while all $\log Z\le 0$ give comparable $\chi^2$ (see Fig. \ref{fig:logzvsincl}). Additional color information can break the degeneracy between atmosphere models. Thus adding \textit{XMM} U data helps resolve this ambiguity.
The best-fit models with the standard $\alpha$ enhancements select slightly sub-solar metallicity with a minimum $\chi^2$/DoF=1.35 for $(\log Z,\alpha)=(-1,0.4)$. $(\log Z, \alpha)=(-0.5,0.2)$ has an increased $\chi^2$; however, the best fit model improves with the limited $(\log Z,\alpha)=(-0.5,0)$ grid. The $(\log Z,\alpha)=(-0.5,0)$ atmosphere represents the global best fit with $\Delta \chi^2 =-1$ improvement over $(\log Z, \alpha)=(-1,0.4)$. Inspecting the synthesized colors, we note that the $\alpha$ changes between $(-0.5,0)$ and $(-0.5,0.2)$ introduce small $\sim 5\%$ differences in the emergent flux, especially in the bluest bands for $T<7000$ K. Unfortunately, the online spectra have insufficient $\alpha$ coverage to fully explore this subtle effect.

\begin{figure*}

    \centering
    \includegraphics[width=0.7\linewidth]{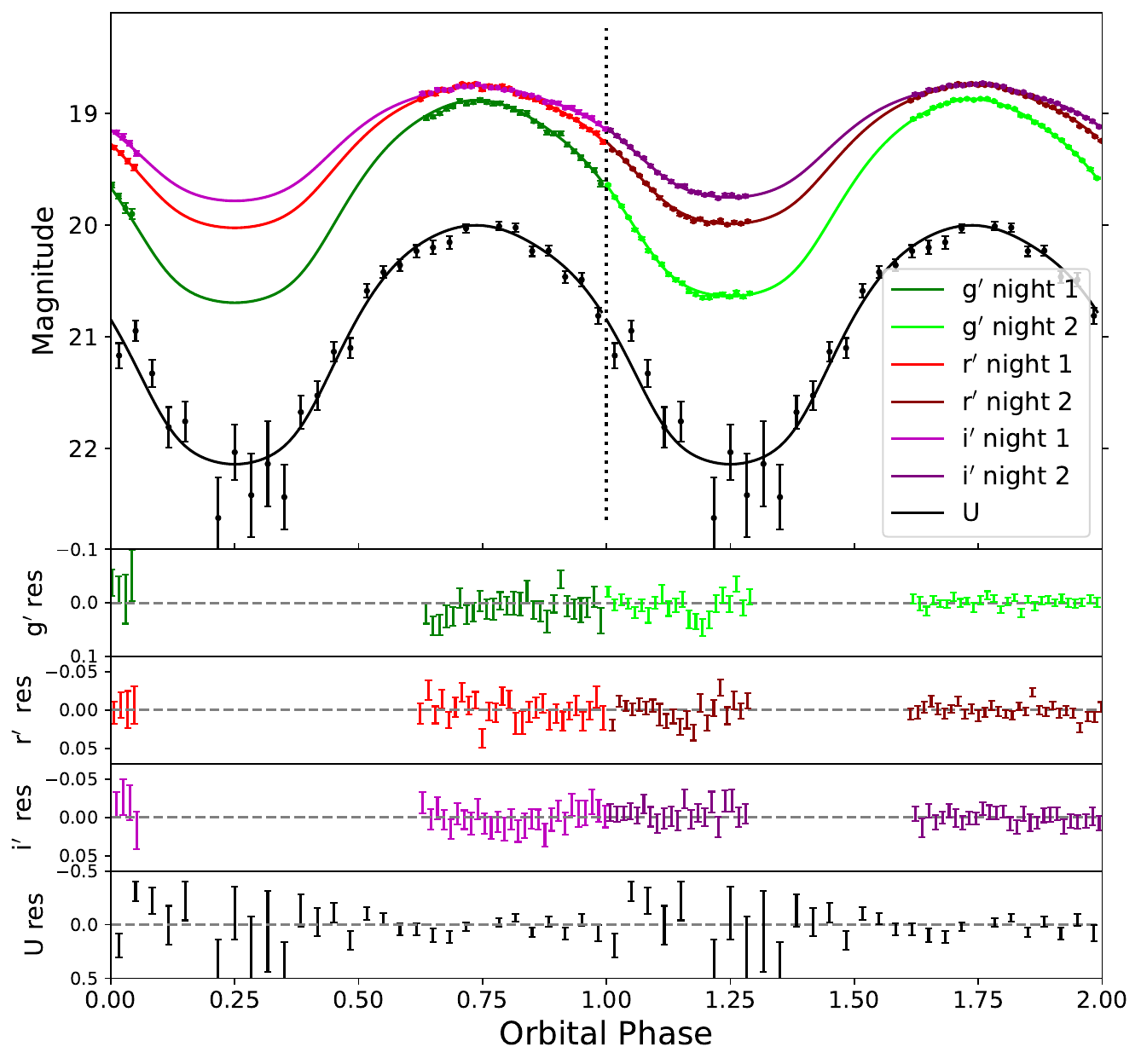}
    \caption{The best-fit companion optical light curve model plotted against the g$^\prime$r$^\prime$i$^\prime$ data from WHT and the OM U data shifted by the estimated calibration offsets along with the residuals. The companion has $(\log Z, \alpha)=(-0.5,0)$. Phase $0\leq\Phi\leq1$ shows WHT night 1 while  $1\leq\Phi\leq2$ shows WHT night 2.}
    \label{fig:OpticalfitLC}
\end{figure*}

The optical spectra of \citet{2015ApJ...809L..10R} also probe the metallicity. In comparing Ca, Mg and Fe absorption line equivalent widths at maximum with BT-Settl model spectra, we see a best match to $\log Z=0$, similar to the spider companions examined in \citet{2019ApJ...883..108D}. The discrepancy might be related to some imprecision in the colors synthesized from the spectra or in the optical broad band flux calibration. In Table \ref{table:2}, we show the estimated parameters of the Ug'r'i' models using the $(\log Z, \alpha)=(-1,0.4)$, $(-0.5,0)$, and $(0,0)$ atmospheres. Fig.\,\ref{fig:OpticalfitLC} shows the best-fit light curve for the $(\log Z,\alpha)=(-0.5,0)$ model.

The larger values of $i$ now preferred imply a lower neutron star mass than found by KR20. To determine this mass (also shown in Table \ref{table:2}), we use the KR20 radial velocity measurement $K_c=429.8\pm3.9$ km/s. Including $1\sigma$ statistical uncertainties, our mass estimates range from $M_{NS}=1.85-2.3$ M$_\odot$, increasing with decreased metallicity. Although deprecated by the photometry, the spectroscopically preferred $\log Z=0$ gives $1.91\pm0.06 M_\odot$ while the photometrically acceptable $\log Z=-1$ fit gives $M_{NS}=2.18\pm0.10$\,M$_\odot$. These are both well less than the ($\log Z=-1$) $M_{NS}=2.24\pm0.09$ M$_\odot$ estimate of KR20.

\begin{table}
\setlength{\tabcolsep}{2pt}
\begin{tabular}{ccccc}

\hline\hline
Parameters  &-1.0/0.4 &   $-0.5/0$ & 0/0&   KR20\\

 \hline
$i$ (deg)  &71.2$^{+2.5}_{-2.0}$  & 77.7$^{+3.7}_{-2.7}$ & 81.5$^{+2.5}_{-2.9}$ &68.9$^{+1.9}_{-1.8}$\\
$M_{NS}$ (M$_\odot$)&2.18$^{+0.10}_{-0.10}$& $1.98^{+0.09}_{-0.08}$& $1.91^{+0.07}_{-0.06}$ &$2.24^{+0.09}_{-0.09}$\\
 $f_c$ & 0.93$^{+0.01}_{-0.01}$& $0.91^{+0.01}_{-0.01}$ & $0.90^{+0.007}_{-0.006}$& 0.94$^{+0.01}_{-0.01}$\\
$L_{H,34}$&$2.51^{+0.10}_{-0.08}$ & $2.51^{+0.09}_{-0.08}$&$2.48^{+0.06}_{-0.04}$&2.6$^{+0.1}_{-0.1}$\\
$T_N$ (K)  &$5590^{+16}_{-17}$ &  $5630^{+15}_{-14}$ & 5731$^{+11}_{-11}$ & 5682$^{+14}_{-15}$\\
$d$ (kpc)  &$3.33^{+0.04}_{-0.04}$ &$3.29^{+0.04}_{-0.04}$&$3.34^{+0.03}_{-0.02}$&$3.30^{+0.04}_{-0.04}$\\
$f_2$ ($\mu$Jy)&$1.7^{+0.2}_{-0.2}$ &  $1.6^{+0.2}_{-0.2}$ & 1.8$^{+0.2}_{-0.2}$&1.2$^{+0.2}_{-0.2}$ \\
$\delta m_r^\dagger$ & $0.027(1)$&  $0.035(1)$ & 0.044(1) & $0.031(1)$\\
$\theta_{HS}$ (deg)& $339.3^{+4.8}_{-4.8}$& $322.6^{+5.9}_{-5.8}$ & $302.2^{+3.9}_{-3.8}$&$324.0^{+10.1}_{-8.7}$\\
$\phi_{HS}$ (deg) &$72.2^{+5.1}_{-8.4}$ &  73.4$^{+4.2}_{-5.6}$ &78.9$^{+2.1}_{-2.8}$& $73.2^{+5.0}_{-8.4}$ \\
$A_{HS}$ &$0.31^{+0.10}_{-0.07}$ & $0.45^{+0.18}_{-0.13}$ &$0.90^{+0.33}_{-0.29}$&$0.6^{+0.4}_{-0.2}$\\
$r_{HS}$ (deg) & 20.8$^{+2.6}_{-3.5}$ &   $16.9^{+3.2}_{-3.1}$ &$17.0^{+2.1}_{-2.1}$  &$15.5^{+3.9}_{-6.8}$\\
$\delta m_U^\dagger$ &  0.472(13)&0.477(13)&0.352(14)& --\\
$\chi^2$/DoF & 351/260 &  349/260 & 388/260 & 297/232 \\
 \hline\hline
\end{tabular}
\caption{Parameter results for optical modeling of WHT+OM J2215 data including the 68\% confidence intervals. We show the best-fitting results with metallicities $(\log Z, \alpha)=(-1,0.4)$, $(-0.5,0)$, and $(0,0)$. $f_c$ is the Roche lobe filling factor, $L_{H,34}$ is the pulsar heating luminosity for an equatorially concentrated heating flux in units of $10^{34}\, {\rm erg\, s^{-1}}$, and $d$ is the source distance. The hot spot coordinates $(\theta_{HS}, \phi_{HS})$ are defined so that $(0, 0)$ refers to the sub-pulsar point at the companion nose. The right column shows the fit results of KR20 which used $\log Z=-1$, assumed a different gravity darkening model, and did not have U data. $^\dagger$ $\delta$ are band offsets with last digit errors. $\delta m_U$ is from the standard SAS flux calibration; however, we see evidence for a systematic calibration shift of $\sim$0.5\,mag.  These fits assume $A_V=0.40$, consistent with KR20, our X-ray fits and 3-D dust maps. We have conducted limited (expensive) tests with free $A_V$ in fits, which shift the solutions to slightly larger $A_V$ with higher $L_H$ and $T_N$ to correct the colors; other parameters remain within $1\sigma$ of the fixed $A_V$ fit values.}  
\label{table:2}
\end{table}

\subsection{X-ray Models}
\begin{figure}
    \centering
    \includegraphics[width=\linewidth]{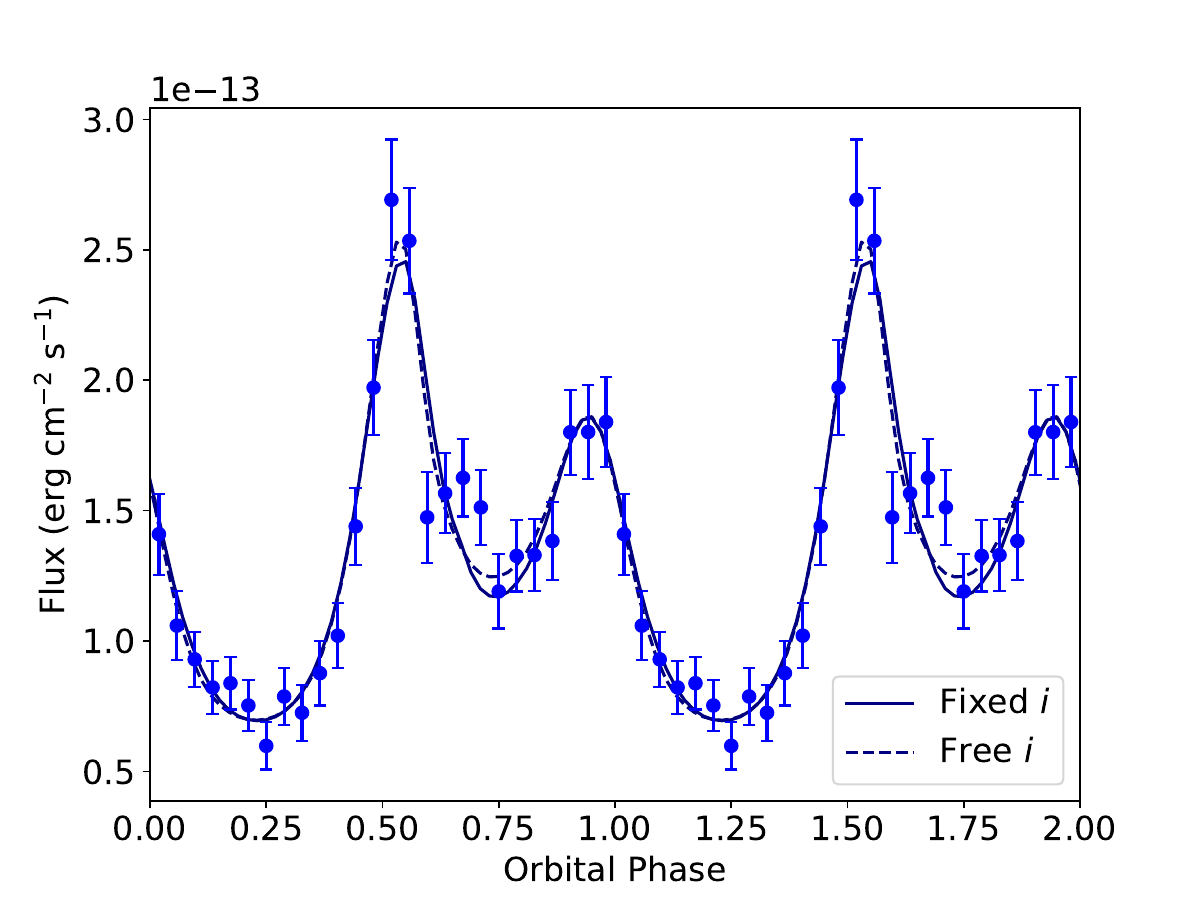}
    \caption{The best-fit IBS X-ray light curves. One model fit fixes $i=80^\circ$, compatible with the optical modeling, while the other model fit leaves $i$ as a free parameter.}
    \label{fig:XrayfitLC}
\end{figure}
\begin{table}
\begin{tabular}{ccc}
 \hline\hline
   Parameters & Fixed $i$ & Free $i$  \\
 \hline
$i$ (deg)  & 80 & 51.8 $^{+16.9}_{-8.5}$  \\
$\beta$  & 2.29$^{+0.24}_{-0.26}$ & 1.52$^{+0.51}_{-0.32}$ \\
$f_v$  & 6.4$^{+5.8}_{-2.6}$ & 3.4$^{+3.2}_{-1.3}$\\
$k$  & 0.076$^{+0.028}_{-0.021}$ & 0.208$^{+0.150}_{-0.105}$   \\
$\delta \Phi$  &0.026$^{+0.008}_{-0.007}$ & 0.032$^{+0.010}_{-0.010}$\\
$\Dot{N}_0$ ($10^{31}$ s$^{-1}$) &  3.7$^{+0.3}_{-0.6}$ &3.2$^{+0.3}_{-0.4}$ \\
$F_b$ ($10^{-14}$ erg cm$^{-2}$ s$^{-1}$) & 4.8$^{+2.0}_{-1.4}$ & 5.0$^{+1.4}_{-1.4}$\\
$\Dot{M}_w$ ($10^{-10}$ $I_{45}$ M$_\odot$ yr$^{-1}$) & 2.0$^{+1.4}_{-1.0}$ & 2.0$^{+1.1}_{-0.9}$\\
$\chi^2$/DoF & 27/20 & 22/19
 \\
 \hline\hline
\end{tabular}
\caption{Parameter results from IBS modeling of the X-ray light curve. $\Dot{M}_w$ is computed assuming $\dot{E}=5.2\times10^{34}$ $I_{45}$ erg s$^{-1}$ \citep{2023ApJ...958..191S}.}
\label{table:3}
\end{table}
We model the X-ray emission with the IBS synchrotron emission prescription of \cite{2019ApJ...879...73K} implemented in the ICARUS IBS code. The code discretizes the IBS (assumed to be thin) into triangular tiles of constant angular size as viewed from the pulsar, representing different zones from which synchrotron radiation is emitted. An electron and positron (hereafter $e^\pm$) population is injected into the IBS with energy spectrum in the flow frame 
\begin{equation}
\label{eq:PL}
    \dot{N}(\gamma_e) d\gamma_e=\dot{N}_0 \gamma_e^{-p} d\gamma_e,
\end{equation} 
where $\gamma_e$ is the particle Lorentz factor in range $\gamma_{min}<\gamma_e<\gamma_{max}$, $\dot{N}_0$ (in e s$^{-1}$) is a global normalization coefficient related to the particle injection rate and $p$ depends on the particle acceleration mechanism.
After injection at a particular tile, the $e^\pm$ population is advected to adjacent tiles with bulk Lorentz factor $\Gamma_{bulk}$. As the particles are advected, synchrotron losses cool the population. The residence time of the particle population in a particular zone is computed in the bulk flow frame as in \cite{2023IBSPolarization}. The synchrotron spectrum at each tile is calculated from the time-averaged particle spectrum in that tile. See \cite{2019ApJ...879...73K} for further details on the emission prescription.

We assume an equatorially concentrated pulsar wind momentum flux $\propto \sin^2\theta_\star$ {in accordance with striped wind models \citep[e.g.][]{1973ApJ...180L.133M}. $\theta_\star$ denotes the angle between the pulsar spin axis (assumed to be aligned with the orbital angular momentum axis since spiders pulsars are strongly recycled) and a position in the wind}.  The pulsar wind collides with a spherical companion wind \citep{2019ApJ...879...73K}. The stellar wind to pulsar wind momentum ratio $\beta=\dot{M_w}v_w c/\dot{E}_{PSR}$ \citep{2016ApJ...828....7R, 2019ApJ...879...73K} governs the overall shock geometry and sets the X-ray peak separation. The relative heights of the peaks are influenced by $i$ as well as the ratio of the stellar wind speed to the orbital speed $f_v=v_w/v_{orb}$. Low values of $f_v$ cause the IBS shock structure to be swept back by the orbital motion, with the geometry tracing out an Archimedean spiral (see \cite{2016ApJ...828....7R} for details on implementation in ICARUS), rather than a symmetric conical structure. The prescription for the bulk Lorentz factor in the post-shock flow is
\begin{equation}
\Gamma_{Bulk}(s)=\Gamma_{nose}\left(1+k\frac{s}{r_0}\right),
\end{equation}
where $s$ is the arclength from the nose to a given point on the IBS, $r_0$ is the nose-standoff distance from the pulsar, $\Gamma_{nose}$ is the bulk Lorentz factor at the nose, and $k$ is a scaling parameter that controls the flow speed increase. The bulk flow direction is assumed to be parallel to the contact discontinuity.

We fit for $\beta$, $f_v$, and $k$. We perform fits with fixed $i=80^\circ$, compatible with the best fitting $(\log Z, \alpha)=(-0.5,0)$ optical model, and with $i$ left as a free parameter.
We fix $\Gamma_{nose}=1.1$, consistent with numerical models \citep{2008MNRAS.387...63B, 2015A&A...581A..27D, 2019ApJ...879...73K}. At the IBS nose, we set $B_{IBS}= 3B_{LC}\left(R_{LC}/r_{0}\right)\approx70$\,G \citep[where {$B_{LC}$ is the magnetic field at the light cylinder radius $R_{LC}$}; the factor of 3 is due to relativistic shock compression;][]{2023IBSPolarization}, based on the inferred orbital separation $a\approx1.5\times10^{11}$ cm, $P_s=2.61$ ms and $\dot{P}_s=2.38\times10^{-20}$ \citep{2023ApJ...958..191S}. We set $p=0.76$ so that our IBS model spectrum has $\Gamma_{IBS}=0.94$ in the $0.5-10$ keV band, consistent with the actual spectral results shown in Table \ref{table:1}. Our fits prefer a forward phase shift  $\delta \Phi\sim0.03$ {from the radio/gamma-ray determined ephemeris and from our optical phase.} We leave $\delta\Phi$ as a fit parameter. We also fit the injected particle spectrum normalization $\dot{N}_0$ and a phase independent flux $F_b$ added as an offset. {We add the offset to account for additional X-ray flux present in the source, possibly due to a pulsar wind nebula (PWN) or the magnetosphere; this component is also present in the spectral fits, as noted above.}

Our best-fit light curves are shown in Fig. \ref{fig:XrayfitLC} and the parameter estimates are shown in Table \ref{table:3}. Our model fits the data moderately well, with $\chi^2/$DoF$=1.35$ when fixing $i$ and $\chi^2/\text{DoF}=1.16$ when leaving $i$ free. This corresponds to an Akaike information criterion (AIC) value of 39 for the fixed $i$ model and 36 for the free $i$ model. Departures from a simple light curve in the phase range $\Phi = 0.6-0.7$ contribute a large factor to the $\chi^2$. In examining the orbit-to-orbit variations, we find that this phase range has somewhat higher variability than the rest of the light curve, but we are unable to attribute the structure in this phase range to IBS flares \citep[as have been seen in other spider pulsars; e.g.][]{2017ApJ...850..100A}. Thus, the flux dip at $\Phi \approx 0.6$ might represent an absorption component, or, alternatively, the flux increase at $\Phi \approx 0.65$ might represent an extra emission component, possibly structure in a thick post-shock flow not captured by our thin-shock geometry. We have left these points in the fit to be conservative; excising these points barely changes the fixed $i$ best-fit parameters. We also note the low bin at $\Phi=0.25$. While at the appropriate phase for a companion eclipse, its significance is low. Furthermore, we would not expect significant occultation of the extended IBS at $i\lesssim 85^\circ$. 

The model fitting gives values $\beta\approx2$ and $f_v\approx3-6$, signaling a rather flat IBS and slow companion wind. If left free, $i$ decreases to {$52^{+17}_{-9} $} deg to accommodate the bridge flux (including the $\Phi \approx 0.65$ component). Such low $i$ implies {$M_{NS}=3.8^{+3.9}_{-1.4}$ M$_\odot$}; the central value is unphysical, although less than $2\sigma$ different from our optical mass determinations. Indeed, $i$ itself is only in $\sim1.5\sigma$ disagreement with the possible optical values. We thus consider the fixed $i$ IBS fit more reliable.

\section{Discussion and Conclusion}
\label{sec:conclusion}
Our optical and X-ray model fits, while imperfect, provide more realistic estimates of the parameters of J2215. Having used the radial velocity $K_{c}$ from  \citet{2020ApJ...892..101K}, we believe that these new values should supersede the results of that paper and those in the less physical modeling of \cite{2018ApJ...859...54L}, especially given the unusual sensitivity to the metallicity. Our best-fit $(\log Z, \alpha)=(-0.5,0)$ result $M_{NS}=1.98\pm0.08$\,M$_\odot$ is more than $2 \sigma$ less than the KR20 estimate. If we adopt $\log Z=0$, as suggested by the optical spectroscopy, the mass lowers by an additional $\sim 1 \sigma$ to $M_{NS}=1.91\pm0.06$\,M$_\odot$. Overall, including the metallicity-induced systematic uncertainty, masses from $\approx 1.85-2.3$\, M$_\odot$ remain acceptable. Improved absolute photometry, especially in UV bands could help break the metallicity degeneracy. Our present result does not strongly constrain the dense matter equation of state given these uncertainties, but can still be valuable for high mass neutron star population studies \citep[e.g.][]{2023PhRvD.108i4014B, 2024PhRvD.109i4030K}.

With well-detected light curve peaks, our X-ray results are much more robust than those of \cite{2016ApJ...828....7R}. Our modeling suggests $\beta\lesssim2.5$ and  $f_v<10$. From these results and $\dot{E}=5.2\times 10^{34}$ $ I_{45}$ erg s$^{-1}$ \citep{2023ApJ...958..191S}, we estimate the companion mass-loss rate
\begin{equation}
    \dot{M}_w=\frac{\beta \dot{E}}{f_v v_{orb} c},
\end{equation}
showing the results in Table \ref{table:3}. $\Dot{M}_w\approx2\times10^{-10}$ $I_{45}$ M$_\odot$ yr$^{-1}$ suggests that the companion should evaporate in $\sim1.5/I_{45}$ Gyr, a modest fraction of the age of the universe, and comparable to the $\tau_c = 1.2$ Gyr characteristic spindown age. This is short enough for J2215 to evolve into an isolated MSP \citep{1988Natur.334..225K}. However, with $\beta$ unusually close to unity, the IBS location and thus mass-loss rate are sensitive to the fit details; these include the small unexplained phase shift.

{The estimated value of the phase-independent flux from the light curve fits $F_b\approx5\times10^{-14}$ erg cm$^{-2}$ s$^{-1}$ is in good agreement with the phase-independent flux estimated from the spectral fitting $F_0$.} These fluxes correspond to a luminosity of $7\times10^{31}$ erg s$^{-1}$  at $d=3.3$ kpc and could originate from magnetospheric emission or, more likely, from a faint, compact PWN \citep[e.g.][]{2008ApJ...682.1166L}. The spectral index $\Gamma_0\approx1.5$ is quite consistent with a PWN source, while the efficiency $L\approx 10^{-3} {\dot E}$ is similar to that of other observed PWNe. The slightly larger value of $F_0$ would include a small amount of emission from the wings of the IBS peaks in the off-phase.

It is also useful to compare our measured energy fluxes with the pulsar spin-down power budget ${\dot E} = 5\times 10^{34}$ $ I_{45}$ $ {\rm erg\,s^{-1}}$. Several neutron star masses have been securely measured above $2$ M$_\odot$, {pointing towards a quite stiff} equation of state. We can estimate $I_{45} \approx [0.8-1.2](M/M_\odot)^{3/2}$ \citep{2005ApJ...629..979L}, i.e. $2.2-3.4$ for $2$ $M_\odot$. The ${\rm sin}^2\theta_\ast$-distributed heating luminosity $L_H \approx 2.5 \times 10^{34}$ $ {\rm erg\, s^{-1}}$ is in good accord with the isotropic GeV gamma-ray luminosity measured by the {\it Fermi} LAT, $L_\gamma = 3.3 \times 10^{34} $ ${\rm erg\, s^{-1}}$, suggesting that our `direct heating' is dominated by the GeV photons. This requires that the radiation be directed toward the companion at the pulsar spin equator, as expected from outer magnetosphere/wind emission models \citep[e.g.][]{2018ApJ...855...94P} for spin-orbit aligned recycled pulsars. This gamma-ray direct heating power is a substantial fraction of the spin-down energy budget, but can be comfortably accommodated for $I_{45}=2-3$. This also reduces our estimated evaporation timescale below 1 Gyr.

Our IBS fit also estimates the $e^\pm/B$ pulsar wind power $\Dot{E}_{pw}$, which is processed by the shock into a power-law electron distribution,
\begin{equation}
    \Dot{E}_{pw}=\frac{\Dot{N}_0}{2-p}m_e c^2(\gamma^{2-p}_{\rm max}-\gamma_{\rm min}^{2-p}).
\end{equation}
The observed 0.5-10 keV X-ray spectrum comes from $10^4< \gamma <10^5$ $e^\pm$ particles. Our fit $\dot{N}_0\sim 3-4\times10^{31}$ s$^{-1}$ thus gives a minimum $\dot{E}_{pw}\gtrsim 5\times10^{31}$ \,erg s$^{-1}$ for $p\approx0.8$. This is well below the available power ${\dot E} - L_H \approx 12 \times 10^{34}$ ${\rm erg\, s^{-1}}$ for $I_{45} \approx 3$. To avoid saturating the residual spin-down power, there is an upper limit of $\gamma_{\rm max} \sim 5.5 \times 10^7$. Higher energy X-ray and gamma-ray measurements can further probe the high $\gamma$ $e^\pm$ population (which cools rapidly). If magnetic reconnection is the primary acceleration mechanism (as suggested by the very hard observed spectra), the maximum electron energy should be $\gamma_{\rm max}\sim\sigma \gamma_{\rm min}$, where $\sigma=B^2/(4\pi \gamma_{\rm min}n_0 m_e c^2)$ is the pulsar wind magnetization and $n_0$ is the e$^\pm$ number density \citep{2011ApJ...741...39S, 2014ApJ...783L..21S}. As expected, the J2215 pulsar wind should be very strongly magnetized with $10\lesssim\sigma\lesssim 6\times10^7$. Finally, the companion hot spot indicates extra localized heating. \citet{2017ApJ...845...42S} attribute companion hot spots to IBS-energized particles precipitating to the companion magnetic poles. In our fits, the small $\beta$ allows the relatively flat IBS shock to capture $\sim40-50\%$ of the $\Dot{E}_{pw}$ flux, while the thermal emission of the hot spot represents $1\times10^{32}$ ${\rm erg\,s^{-1}}$. Only a small fraction $\sim 2 \times 10^{-3}$ of the maximum available IBS-processed pulsar wind power needs to reach the surface to heat the companion magnetic pole. Simultaneous optical/UV observations would be useful to re-check the U band normalization and further constrain the hot spot contribution and atmosphere metallicity. 
\bigskip

J2215's high inferred mass-loss rate makes this system a plausible isolated MSP progenitor, while its unusually flat IBS may reprocess a large fraction of the spin-down power. The hard X-ray spectrum and sharp peaks provide an important probe of relativistic shock dynamics and particle acceleration. Higher energy observations (e.g. with NuSTAR) should further constrain the IBS $e^\pm$ population. Since the prominent IBS peaks indicate beamed emission and thus moderate bulk Lorentz factors in the IBS flow, the various orbital phases probe different portions of the shock with varying obliquity. Thus, IBS-dominated spider pulsars, like J2215, show emission from a range of relativistic shock geometries. Future phase-resolved spectrum and polarization measurements of these natural laboratories can be compared with numerical simulations to advance our understanding of oblique, strongly magnetized relativistic shocks. 

\section*{Acknowledgements}
The authors thank Dinesh Kandel for help using ICARUS and for clarifications on the previous optical modeling as well as the anonymous referee for helpful feedback. This work was supported in part by NASA grant 80NSSC22K1506.
A.S. acknowledges the support of the Stanford University Physics Department Fellowship and the National Science Foundation Graduate Research Fellowship.
\bibliography{refs}{}

\begin{thebibliography}{}
\expandafter\ifx\csname natexlab\endcsname\relax\def\natexlab#1{#1}\fi
\providecommand{\url}[1]{\href{#1}{#1}}
\providecommand{\dodoi}[1]{doi:~\href{http://doi.org/#1}{\nolinkurl{#1}}}
\providecommand{\doeprint}[1]{\href{http://ascl.net/#1}{\nolinkurl{http://ascl.net/#1}}}
\providecommand{\doarXiv}[1]{\href{https://arxiv.org/abs/#1}{\nolinkurl{https://arxiv.org/abs/#1}}}

\bibitem[{{Allard}(2014)}]{2014IAUS..299..271A}
{Allard}, F. 2014, in Exploring the Formation and Evolution of Planetary
  Systems, ed. M.~{Booth}, B.~C. {Matthews}, \& J.~R. {Graham}, Vol. 299,
  271--272, \dodoi{10.1017/S1743921313008545}

\bibitem[{{An} {et~al.}(2017){An}, {Romani}, {Johnson}, {Kerr}, \&
  {Clark}}]{2017ApJ...850..100A}
{An}, H., {Romani}, R.~W., {Johnson}, T., {Kerr}, M., \& {Clark}, C.~J. 2017,
  \apj, 850, 100, \dodoi{10.3847/1538-4357/aa947f}

\bibitem[{{Anders} \& {Grevesse}(1989)}]{1989GeCoA..53..197A}
{Anders}, E., \& {Grevesse}, N. 1989, \gca, 53, 197,
  \dodoi{10.1016/0016-7037(89)90286-X}

\bibitem[{{Atwood} {et~al.}(2009){Atwood}, {Abdo}, {Ackermann}, {Althouse},
  {Anderson}, {Axelsson}, {Baldini}, {Ballet}, {Band}, {Barbiellini},
  {Bartelt}, {Bastieri}, {Baughman}, {Bechtol}, {B{\'e}d{\'e}r{\`e}de},
  {Bellardi}, {Bellazzini}, {Berenji}, {Bignami}, {Bisello}, {Bissaldi},
  {Blandford}, {Bloom}, {Bogart}, {Bonamente}, {Bonnell}, {Borgland},
  {Bouvier}, {Bregeon}, {Brez}, {Brigida}, {Bruel}, {Burnett}, {Busetto},
  {Caliandro}, {Cameron}, {Caraveo}, {Carius}, {Carlson}, {Casandjian},
  {Cavazzuti}, {Ceccanti}, {Cecchi}, {Charles}, {Chekhtman}, {Cheung},
  {Chiang}, {Chipaux}, {Cillis}, {Ciprini}, {Claus}, {Cohen-Tanugi},
  {Condamoor}, {Conrad}, {Corbet}, {Corucci}, {Costamante}, {Cutini}, {Davis},
  {Decotigny}, {DeKlotz}, {Dermer}, {de Angelis}, {Digel}, {do Couto e Silva},
  {Drell}, {Dubois}, {Dumora}, {Edmonds}, {Fabiani}, {Farnier}, {Favuzzi},
  {Flath}, {Fleury}, {Focke}, {Funk}, {Fusco}, {Gargano}, {Gasparrini},
  {Gehrels}, {Gentit}, {Germani}, {Giebels}, {Giglietto}, {Giommi}, {Giordano},
  {Glanzman}, {Godfrey}, {Grenier}, {Grondin}, {Grove}, {Guillemot}, {Guiriec},
  {Haller}, {Harding}, {Hart}, {Hays}, {Healey}, {Hirayama}, {Hjalmarsdotter},
  {Horn}, {Hughes}, {J{\'o}hannesson}, {Johansson}, {Johnson}, {Johnson},
  {Johnson}, {Johnson}, {Kamae}, {Katagiri}, {Kataoka}, {Kavelaars}, {Kawai},
  {Kelly}, {Kerr}, {Klamra}, {Kn{\"o}dlseder}, {Kocian}, {Komin}, {Kuehn},
  {Kuss}, {Landriu}, {Latronico}, {Lee}, {Lee}, {Lemoine-Goumard}, {Lionetto},
  {Longo}, {Loparco}, {Lott}, {Lovellette}, {Lubrano}, {Madejski}, {Makeev},
  {Marangelli}, {Massai}, {Mazziotta}, {McEnery}, {Menon}, {Meurer},
  {Michelson}, {Minuti}, {Mirizzi}, {Mitthumsiri}, {Mizuno}, {Moiseev},
  {Monte}, {Monzani}, {Moretti}, {Morselli}, {Moskalenko}, {Murgia},
  {Nakamori}, {Nishino}, {Nolan}, {Norris}, {Nuss}, {Ohno}, {Ohsugi}, {Omodei},
  {Orlando}, {Ormes}, {Paccagnella}, {Paneque}, {Panetta}, {Parent}, {Pearce},
  {Pepe}, {Perazzo}, {Pesce-Rollins}, {Picozza}, {Pieri}, {Pinchera}, {Piron},
  {Porter}, {Poupard}, {Rain{\`o}}, {Rando}, {Rapposelli}, {Razzano}, {Reimer},
  {Reimer}, {Reposeur}, {Reyes}, {Ritz}, {Rochester}, {Rodriguez}, {Romani},
  {Roth}, {Russell}, {Ryde}, {Sabatini}, {Sadrozinski}, {Sanchez}, {Sander},
  {Sapozhnikov}, {Parkinson}, {Scargle}, {Schalk}, {Scolieri}, {Sgr{\`o}},
  {Share}, {Shaw}, {Shimokawabe}, {Shrader}, {Sierpowska-Bartosik}, {Siskind},
  {Smith}, {Smith}, {Spandre}, {Spinelli}, {Starck}, {Stephens}, {Strickman},
  {Strong}, {Suson}, {Tajima}, {Takahashi}, {Takahashi}, {Tanaka}, {Tenze},
  {Tether}, {Thayer}, {Thayer}, {Thompson}, {Tibaldo}, {Tibolla}, {Torres},
  {Tosti}, {Tramacere}, {Turri}, {Usher}, {Vilchez}, {Vitale}, {Wang},
  {Watters}, {Winer}, {Wood}, {Ylinen}, \& {Ziegler}}]{2009ApJ...697.1071A}
{Atwood}, W.~B., {Abdo}, A.~A., {Ackermann}, M., {et~al.} 2009, \apj, 697,
  1071, \dodoi{10.1088/0004-637X/697/2/1071}

\bibitem[{{Balucinska-Church} \& {McCammon}(1992)}]{1992ApJ...400..699B}
{Balucinska-Church}, M., \& {McCammon}, D. 1992, \apj, 400, 699,
  \dodoi{10.1086/172032}

\bibitem[{{Bogovalov} {et~al.}(2008){Bogovalov}, {Khangulyan}, {Koldoba},
  {Ustyugova}, \& {Aharonian}}]{2008MNRAS.387...63B}
{Bogovalov}, S.~V., {Khangulyan}, D.~V., {Koldoba}, A.~V., {Ustyugova}, G.~V.,
  \& {Aharonian}, F.~A. 2008, \mnras, 387, 63,
  \dodoi{10.1111/j.1365-2966.2008.13226.x}

\bibitem[{{Brandes} {et~al.}(2023){Brandes}, {Weise}, \&
  {Kaiser}}]{2023PhRvD.108i4014B}
{Brandes}, L., {Weise}, W., \& {Kaiser}, N. 2023, \prd, 108, 094014,
  \dodoi{10.1103/PhysRevD.108.094014}

\bibitem[{{Breton} {et~al.}(2013){Breton}, {van Kerkwijk}, {Roberts},
  {Hessels}, {Camilo}, {McLaughlin}, {Ransom}, {Ray}, \&
  {Stairs}}]{2013ApJ...769..108B}
{Breton}, R.~P., {van Kerkwijk}, M.~H., {Roberts}, M.~S.~E., {et~al.} 2013,
  \apj, 769, 108, \dodoi{10.1088/0004-637X/769/2/108}

\bibitem[{{Broderick} {et~al.}(2016){Broderick}, {Fender}, {Breton}, {Stewart},
  {Rowlinson}, {Swinbank}, {Hessels}, {Staley}, {van der Horst}, {Bell},
  {Carbone}, {Cendes}, {Corbel}, {Eisl{\"o}ffel}, {Falcke}, {Grie{\ss}meier},
  {Hassall}, {Jonker}, {Kramer}, {Kuniyoshi}, {Law}, {Markoff}, {Molenaar},
  {Pietka}, {Scheers}, {Serylak}, {Stappers}, {ter Veen}, {van Leeuwen},
  {Wijers}, {Wijnands}, {Wise}, \& {Zarka}}]{2016MNRAS.459.2681B}
{Broderick}, J.~W., {Fender}, R.~P., {Breton}, R.~P., {et~al.} 2016, \mnras,
  459, 2681, \dodoi{10.1093/mnras/stw794}

\bibitem[{{Buchner} {et~al.}(2014){Buchner}, {Georgakakis}, {Nandra}, {Hsu},
  {Rangel}, {Brightman}, {Merloni}, {Salvato}, {Donley}, \&
  {Kocevski}}]{2014A&A...564A.125B}
{Buchner}, J., {Georgakakis}, A., {Nandra}, K., {et~al.} 2014, \aap, 564, A125,
  \dodoi{10.1051/0004-6361/201322971}

\bibitem[{{Claret} \& {Bloemen}(2011)}]{2011A&A...529A..75C}
{Claret}, A., \& {Bloemen}, S. 2011, \aap, 529, A75,
  \dodoi{10.1051/0004-6361/201116451}

\bibitem[{{Doe} {et~al.}(2007){Doe}, {Nguyen}, {Stawarz}, {Refsdal},
  {Siemiginowska}, {Burke}, {Evans}, {Evans}, {McDowell}, {Houck}, \&
  {Nowak}}]{2007ASPC..376..543D}
{Doe}, S., {Nguyen}, D., {Stawarz}, C., {et~al.} 2007, in Astronomical Society
  of the Pacific Conference Series, Vol. 376, Astronomical Data Analysis
  Software and Systems XVI, ed. R.~A. {Shaw}, F.~{Hill}, \& D.~J. {Bell}, 543

\bibitem[{{Draghis} {et~al.}(2019){Draghis}, {Romani}, {Filippenko}, {Brink},
  {Zheng}, {Halpern}, \& {Camilo}}]{2019ApJ...883..108D}
{Draghis}, P., {Romani}, R.~W., {Filippenko}, A.~V., {et~al.} 2019, \apj, 883,
  108, \dodoi{10.3847/1538-4357/ab378b}

\bibitem[{{Dubus} {et~al.}(2015){Dubus}, {Lamberts}, \&
  {Fromang}}]{2015A&A...581A..27D}
{Dubus}, G., {Lamberts}, A., \& {Fromang}, S. 2015, \aap, 581, A27,
  \dodoi{10.1051/0004-6361/201425394}

\bibitem[{{Feroz} \& {Hobson}(2008)}]{2008MNRAS.384..449F}
{Feroz}, F., \& {Hobson}, M.~P. 2008, \mnras, 384, 449,
  \dodoi{10.1111/j.1365-2966.2007.12353.x}

\bibitem[{{Feroz} {et~al.}(2009){Feroz}, {Hobson}, \&
  {Bridges}}]{2009MNRAS.398.1601F}
{Feroz}, F., {Hobson}, M.~P., \& {Bridges}, M. 2009, \mnras, 398, 1601,
  \dodoi{10.1111/j.1365-2966.2009.14548.x}

\bibitem[{{Feroz} {et~al.}(2019){Feroz}, {Hobson}, {Cameron}, \&
  {Pettitt}}]{2019OJAp....2E..10F}
{Feroz}, F., {Hobson}, M.~P., {Cameron}, E., \& {Pettitt}, A.~N. 2019, The Open
  Journal of Astrophysics, 2, 10, \dodoi{10.21105/astro.1306.2144}

\bibitem[{{Freeman} {et~al.}(2001){Freeman}, {Doe}, \&
  {Siemiginowska}}]{2001SPIE.4477...76F}
{Freeman}, P., {Doe}, S., \& {Siemiginowska}, A. 2001, in Society of
  Photo-Optical Instrumentation Engineers (SPIE) Conference Series, Vol. 4477,
  Astronomical Data Analysis, ed. J.-L. {Starck} \& F.~D. {Murtagh}, 76--87,
  \dodoi{10.1117/12.447161}

\bibitem[{{Gentile} {et~al.}(2014){Gentile}, {Roberts}, {McLaughlin}, {Camilo},
  {Hessels}, {Kerr}, {Ransom}, {Ray}, \& {Stairs}}]{2014ApJ...783...69G}
{Gentile}, P.~A., {Roberts}, M.~S.~E., {McLaughlin}, M.~A., {et~al.} 2014,
  \apj, 783, 69, \dodoi{10.1088/0004-637X/783/2/69}

\bibitem[{{Green} {et~al.}(2018){Green}, {Schlafly}, {Finkbeiner}, {Rix},
  {Martin}, {Burgett}, {Draper}, {Flewelling}, {Hodapp}, {Kaiser}, {Kudritzki},
  {Magnier}, {Metcalfe}, {Tonry}, {Wainscoat}, \&
  {Waters}}]{2018MNRAS.478..651G}
{Green}, G.~M., {Schlafly}, E.~F., {Finkbeiner}, D., {et~al.} 2018, \mnras,
  478, 651, \dodoi{10.1093/mnras/sty1008}

\bibitem[{{Hessels} {et~al.}(2011){Hessels}, {Roberts}, {McLaughlin}, {Ray},
  {Bangale}, {Ransom}, {Kerr}, {Camilo}, \& {Decesar}}]{2011AIPC.1357...40H}
{Hessels}, J.~W.~T., {Roberts}, M.~S.~E., {McLaughlin}, M.~A., {et~al.} 2011,
  in American Institute of Physics Conference Series, Vol. 1357, Radio Pulsars:
  An Astrophysical Key to Unlock the Secrets of the Universe, ed. M.~{Burgay},
  N.~{D'Amico}, P.~{Esposito}, A.~{Pellizzoni}, \& A.~{Possenti}, 40--43,
  \dodoi{10.1063/1.3615072}

\bibitem[{{Hui} \& {Li}(2019)}]{2019Galax...7...93H}
{Hui}, C.~Y., \& {Li}, K.~L. 2019, Galaxies, 7, 93,
  \dodoi{10.3390/galaxies7040093}

\bibitem[{{Jansen} {et~al.}(2001){Jansen}, {Lumb}, {Altieri}, {Clavel}, {Ehle},
  {Erd}, {Gabriel}, {Guainazzi}, {Gondoin}, {Much}, {Munoz}, {Santos},
  {Schartel}, {Texier}, \& {Vacanti}}]{2001A&A...365L...1J}
{Jansen}, F., {Lumb}, D., {Altieri}, B., {et~al.} 2001, \aap, 365, L1,
  \dodoi{10.1051/0004-6361:20000036}

\bibitem[{{Kandel} \& {Romani}(2020)}]{2020ApJ...892..101K}
{Kandel}, D., \& {Romani}, R.~W. 2020, \apj, 892, 101,
  \dodoi{10.3847/1538-4357/ab7b62}

\bibitem[{{Kandel} \& {Romani}(2023)}]{2023ApJ...942....6K}
---. 2023, \apj, 942, 6, \dodoi{10.3847/1538-4357/aca524}

\bibitem[{{Kandel} {et~al.}(2019){Kandel}, {Romani}, \&
  {An}}]{2019ApJ...879...73K}
{Kandel}, D., {Romani}, R.~W., \& {An}, H. 2019, \apj, 879, 73,
  \dodoi{10.3847/1538-4357/ab24d9}

\bibitem[{{Kandel} {et~al.}(2021){Kandel}, {Romani}, \&
  {An}}]{2021ApJ...917L..13K}
---. 2021, \apjl, 917, L13, \dodoi{10.3847/2041-8213/ac15f7}

\bibitem[{{Kluzniak} {et~al.}(1988){Kluzniak}, {Ruderman}, {Shaham}, \&
  {Tavani}}]{1988Natur.334..225K}
{Kluzniak}, W., {Ruderman}, M., {Shaham}, J., \& {Tavani}, M. 1988, \nat, 334,
  225, \dodoi{10.1038/334225a0}

\bibitem[{{Komoltsev} {et~al.}(2024){Komoltsev}, {Somasundaram}, {Gorda},
  {Kurkela}, {Margueron}, \& {Tews}}]{2024PhRvD.109i4030K}
{Komoltsev}, O., {Somasundaram}, R., {Gorda}, T., {et~al.} 2024, \prd, 109,
  094030, \dodoi{10.1103/PhysRevD.109.094030}

\bibitem[{{Lattimer} \& {Prakash}(2007)}]{2007PhR...442..109L}
{Lattimer}, J.~M., \& {Prakash}, M. 2007, \physrep, 442, 109,
  \dodoi{10.1016/j.physrep.2007.02.003}

\bibitem[{{Lattimer} \& {Schutz}(2005)}]{2005ApJ...629..979L}
{Lattimer}, J.~M., \& {Schutz}, B.~F. 2005, \apj, 629, 979,
  \dodoi{10.1086/431543}

\bibitem[{{Li} {et~al.}(2008){Li}, {Lu}, \& {Li}}]{2008ApJ...682.1166L}
{Li}, X.-H., {Lu}, F.-J., \& {Li}, Z. 2008, \apj, 682, 1166,
  \dodoi{10.1086/589495}

\bibitem[{{Linares}(2014)}]{2014ApJ...795...72L}
{Linares}, M. 2014, \apj, 795, 72, \dodoi{10.1088/0004-637X/795/1/72}

\bibitem[{{Linares} {et~al.}(2018){Linares}, {Shahbaz}, \&
  {Casares}}]{2018ApJ...859...54L}
{Linares}, M., {Shahbaz}, T., \& {Casares}, J. 2018, \apj, 859, 54,
  \dodoi{10.3847/1538-4357/aabde6}

\bibitem[{{Michel}(1973)}]{1973ApJ...180L.133M}
{Michel}, F.~C. 1973, \apjl, 180, L133, \dodoi{10.1086/181169}

\bibitem[{{Philippov} \& {Spitkovsky}(2018)}]{2018ApJ...855...94P}
{Philippov}, A.~A., \& {Spitkovsky}, A. 2018, \apj, 855, 94,
  \dodoi{10.3847/1538-4357/aaabbc}

\bibitem[{{Ray} {et~al.}(2012){Ray}, {Abdo}, {Parent}, {Bhattacharya},
  {Bhattacharyya}, {Camilo}, {Cognard}, {Theureau}, {Ferrara}, {Harding},
  {Thompson}, {Freire}, {Guillemot}, {Gupta}, {Roy}, {Hessels}, {Johnston},
  {Keith}, {Shannon}, {Kerr}, {Michelson}, {Romani}, {Kramer}, {McLaughlin},
  {Ransom}, {Roberts}, {Saz Parkinson}, {Ziegler}, {Smith}, {Stappers},
  {Weltevrede}, \& {Wood}}]{2012arXiv1205.3089R}
{Ray}, P.~S., {Abdo}, A.~A., {Parent}, D., {et~al.} 2012, arXiv e-prints,
  arXiv:1205.3089, \dodoi{10.48550/arXiv.1205.3089}

\bibitem[{{Roberts}(2013)}]{2013IAUS..291..127R}
{Roberts}, M. S.~E. 2013, in Neutron Stars and Pulsars: Challenges and
  Opportunities after 80 years, ed. J.~{van Leeuwen}, Vol. 291, 127--132,
  \dodoi{10.1017/S174392131202337X}

\bibitem[{{Romani} {et~al.}(2015){Romani}, {Graham}, {Filippenko}, \&
  {Kerr}}]{2015ApJ...809L..10R}
{Romani}, R.~W., {Graham}, M.~L., {Filippenko}, A.~V., \& {Kerr}, M. 2015,
  \apjl, 809, L10, \dodoi{10.1088/2041-8205/809/1/L10}

\bibitem[{{Romani} \& {Sanchez}(2016)}]{2016ApJ...828....7R}
{Romani}, R.~W., \& {Sanchez}, N. 2016, \apj, 828, 7,
  \dodoi{10.3847/0004-637X/828/1/7}

\bibitem[{{Sanchez} \& {Romani}(2017)}]{2017ApJ...845...42S}
{Sanchez}, N., \& {Romani}, R.~W. 2017, \apj, 845, 42,
  \dodoi{10.3847/1538-4357/aa7a02}

\bibitem[{{Schroeder} \& {Halpern}(2014)}]{2014ApJ...793...78S}
{Schroeder}, J., \& {Halpern}, J. 2014, \apj, 793, 78,
  \dodoi{10.1088/0004-637X/793/2/78}

\bibitem[{{Sironi} \& {Spitkovsky}(2011)}]{2011ApJ...741...39S}
{Sironi}, L., \& {Spitkovsky}, A. 2011, \apj, 741, 39,
  \dodoi{10.1088/0004-637X/741/1/39}

\bibitem[{{Sironi} \& {Spitkovsky}(2014)}]{2014ApJ...783L..21S}
---. 2014, \apjl, 783, L21, \dodoi{10.1088/2041-8205/783/1/L21}

\bibitem[{{Smith} {et~al.}(2023){Smith}, {Abdollahi}, {Ajello}, {Bailes},
  {Baldini}, {Ballet}, {Baring}, {Bassa}, {Becerra Gonzalez}, {Bellazzini},
  {Berretta}, {Bhattacharyya}, {Bissaldi}, {Bonino}, {Bottacini}, {Bregeon},
  {Bruel}, {Burgay}, {Burnett}, {Cameron}, {Camilo}, {Caputo}, {Caraveo},
  {Cavazzuti}, {Chiaro}, {Ciprini}, {Clark}, {Cognard}, {Corongiu},
  {Cristarella Orestano}, {Crnogorcevic}, {Cuoco}, {Cutini}, {D'Ammando}, {de
  Angelis}, {DeCesar}, {De Gaetano}, {de Menezes}, {Deneva}, {de Palma}, {Di
  Lalla}, {Dirirsa}, {Di Venere}, {Dom{\'\i}nguez}, {Dumora}, {Fegan},
  {Ferrara}, {Fiori}, {Fleischhack}, {Flynn}, {Franckowiak}, {Freire},
  {Fukazawa}, {Fusco}, {Galanti}, {Gammaldi}, {Gargano}, {Gasparrini},
  {Giacchino}, {Giglietto}, {Giordano}, {Giroletti}, {Green}, {Grenier},
  {Guillemot}, {Guiriec}, {Gustafsson}, {Harding}, {Hays}, {Hewitt}, {Horan},
  {Hou}, {Jankowski}, {Johnson}, {Johnson}, {Johnston}, {Kataoka}, {Keith},
  {Kerr}, {Kramer}, {Kuss}, {Latronico}, {Lee}, {Li}, {Li}, {Limyansky},
  {Longo}, {Loparco}, {Lorusso}, {Lovellette}, {Lower}, {Lubrano}, {Lyne},
  {Maan}, {Maldera}, {Manchester}, {Manfreda}, {Marelli}, {Mart{\'\i}-Devesa},
  {Mazziotta}, {McEnery}, {Mereu}, {Michelson}, {Mickaliger}, {Mitthumsiri},
  {Mizuno}, {Moiseev}, {Monzani}, {Morselli}, {Negro}, {Nemmen}, {Nieder},
  {Nuss}, {Omodei}, {Orienti}, {Orlando}, {Ormes}, {Palatiello}, {Paneque},
  {Panzarini}, {Parthasarathy}, {Persic}, {Pesce-Rollins}, {Pillera}, {Poon},
  {Porter}, {Possenti}, {Principe}, {Rain{\`o}}, {Rando}, {Ransom}, {Ray},
  {Razzano}, {Razzaque}, {Reimer}, {Reimer}, {Renault-Tinacci}, {Romani},
  {S{\'a}nchez-Conde}, {Saz Parkinson}, {Scotton}, {Serini}, {Sgr{\`o}},
  {Shannon}, {Sharma}, {Shen}, {Siskind}, {Spandre}, {Spinelli}, {Stappers},
  {Stephens}, {Suson}, {Tabassum}, {Tajima}, {Tak}, {Theureau}, {Thompson},
  {Tibolla}, {Torres}, {Valverde}, {Venter}, {Wadiasingh}, {Wang}, {Wang},
  {Wang}, {Weltevrede}, {Wood}, {Yan}, {Zaharijas}, {Zhang}, \&
  {Zhu}}]{2023ApJ...958..191S}
{Smith}, D.~A., {Abdollahi}, S., {Ajello}, M., {et~al.} 2023, \apj, 958, 191,
  \dodoi{10.3847/1538-4357/acee67}

\bibitem[{{Steiner} {et~al.}(2013){Steiner}, {Lattimer}, \&
  {Brown}}]{2013ApJ...765L...5S}
{Steiner}, A.~W., {Lattimer}, J.~M., \& {Brown}, E.~F. 2013, \apjl, 765, L5,
  \dodoi{10.1088/2041-8205/765/1/L5}

\bibitem[{{Sullivan} \& {Romani}(2023)}]{2023IBSPolarization}
{Sullivan}, A.~G., \& {Romani}, R.~W. 2023, \apj, 959, 81,
  \dodoi{10.3847/1538-4357/ad09ae}

\bibitem[{{van Paradijs} {et~al.}(1988){van Paradijs}, {Allington-Smith},
  {Callanan}, {Charles}, {Hassall}, {Machin}, {Mason}, {Naylor}, \&
  {Smale}}]{1988Natur.334..684V}
{van Paradijs}, J., {Allington-Smith}, J., {Callanan}, P., {et~al.} 1988, \nat,
  334, 684, \dodoi{10.1038/334684a0}

\bibitem[{{Wadiasingh} {et~al.}(2017){Wadiasingh}, {Harding}, {Venter},
  {B{\"o}ttcher}, \& {Baring}}]{2017ApJ...839...80W}
{Wadiasingh}, Z., {Harding}, A.~K., {Venter}, C., {B{\"o}ttcher}, M., \&
  {Baring}, M.~G. 2017, \apj, 839, 80, \dodoi{10.3847/1538-4357/aa69bf}

\bibitem[{{Yao} {et~al.}(2017){Yao}, {Manchester}, \&
  {Wang}}]{2017ApJ...835...29Y}
{Yao}, J.~M., {Manchester}, R.~N., \& {Wang}, N. 2017, \apj, 835, 29,
  \dodoi{10.3847/1538-4357/835/1/29}

\end{thebibliography}
\bibliographystyle{aasjournal}



\end{document}